# Conditional Response Probabilities Confirm Two Stages of Free Recall, Working Memory Very Localized, Second Stage Delocalized, Both with Strong Tendencies to Forward Subsequent Recalls

Eugen Tarnow

18-11 Radburn Road

Fair Lawn, NJ 07410

etarnow@avabiz.com

## Abstract:

Recently it was shown that free recall consists of two stages: the first few recalls empty working memory and a second stage concludes the recall (Tarnow, 2015; for a review of the theoretical prediction see Murdock, 1974). Here I investigate conditional response probabilities in Murdock's 40-1 (1962) free recall dataset. I find that the conditional response probabilities confirm the presence of two stages. The first stage is characterized by a large enhancement of the forward subsequent recall (up to 30 times chance) and a large suppression of backward recalls (up to 26 times smaller than chance for recalls between 10 and 40 items away). The second stage forward subsequent recall is enhanced by a factor of 5 and the probability of backward and forward recalls are concentrated at small distances. Thus both stages favor forward subsequent recalls. As in Tarnow (2015), the first stage is much more localized than the second stage. Working memory capacity is estimated in six ways to be about 4.5 word items for the 40-1 word list.

## Introduction

Free recall stands out as one of the great unsolved mysteries of modern psychology (for reviews, please see, for example, Watkins, 1974; Murdock, 1974; Laming, 2010). Items in a list are displayed or read to subjects who are then asked to retrieve the items. The results (for example, Murdock, 1960; Murdock, 1962; Murdock, 1974) have defied explanation. Why do we remember primarily items in the beginning and in the end of the list, but not items in the middle, creating the famous u-shaped curve of probability of recall versus serial position? Why can we remember 50-100 items in cued recall but only 6-8 items in free recall?

Some of the mystery has been removed. We now know explicitly that free recall consists of two stages (Tarnow, 2015; for a review of the experiments and theory which predicted the two stages see Murdock, 1974). In the first stage working memory is emptied and in the second stage a different retrieval process occurs. Working memory is responsible for the recency part of the serial position curve and for some of the first item recall for short lists (Tarnow, 2015).

From Tarnow (2015): In Fig. 1 is shown the recall distributions of recalls 1-8 from the 10-2 free recall dataset of Murdock (1962). These distributions show direct evidence for a two stage process. By definition the first recall comes from working memory, and from the similarity of the $2^{nd}$ and $3^{rd}$ recalls these also come from working memory. The last three recalls come from a second stage of recall and are similar to each other but differ from the first stage. Recalls 4 and 5 are a combination of the two. In each recall is plotted a best linear fit which expresses the balance between recency (positive slope) and primacy (negative slope). As we see the slopes go from primacy for the emptying of working memory to recency for the second stage. Working memory can be seen as responsible for recency (consistent with previous work, see Watkins, 1974 and Glanzer, 1982); primacy comes from the secondary process though working memory adds the first items in the shorter 10-2 list; together they create a u-shaped serial position curve.

From Tarnow (2015): The slopes as a function of recall are plotted in Fig. 2. The curve is a smoothed step function. The middle of the step function "discontinuity" corresponds to the capacity of working memory and is 4 for the 10-2 data.

In this contribution I will investigate how the conditional response probability is impacted by the recall stages.

## Method

This article makes use of the Murdock (1962) data set (downloaded from the Computational Memory Lab at the University of Pennsylvania (http://memory.psych.upenn.edu/DataArchive)). In Table 1 is summarized the experimental process which generated the data set used in this paper.

## Results & Discussion

Recent findings about conditional response probabilities include observations that conditional response probabilities are rather complex (Farrell & Lewandowsky, 2008) and strongly influenced by the boundary conditions – that there is a beginning and an end to a free recall list (Tarnow, 2015).

In this paper we will study conditional response probabilities for the 40-1 experiment. This is the experiment with the longest word list for which boundary conditions should matter the least. To better understand the data, we will start with a simplistic random model and gradually refine it with real data. First, the response probability curve from chance recalls is displayed in Fig. 3. We have assumed that all items are equally likely to be recalled and equally likely to be followed by any other item (for simplicity even disregarding the condition that an item can only be recalled once). The chance conditional response probabilities for large distances are small and increase linearly towards small distances. The probability of a very small distance is $1/N$ where $N$ is the list length, and 2.5% for the 40-1 list (it would be a larger 10% for a ten item list). If the boundary conditions did not matter, each distance would have an equally likely probability of $1/(2N) = 1.25\%$ for the 40-1 list.

Second, we add historical recall information and consider the conditional response probability curve of a single random recall given the recall history up to that point (Fig. 4). Since the last "random" recall probes the structure of the previous recall history we can distinguish the recall stages (Tarnow, 2015). The first stage of free recall, the emptying of working memory localized at recency boundary, makes a random conditional response appear as a "square wave" curve. The second stage of free recall, delocalized, makes the conditional response curve appear random as in Fig. 3; intermediate recalls are combinations of the two shapes. The peaks of these chance recalls are all at about 2.5%. The halfway point between the two stages is around recall 4.5 which is a first estimate of the capacity of working memory.

Third, the actual conditional response probabilities are displayed in Fig. 5. Note the large difference in scale from Figs. 3 and 4: instead of a peak at 2.5%, the peak reaches 60% for the beginning of the first stage and is gradually lowered to about 15% for the second stage (the result for the first conditional recall is similar to Farrell & Lewandowsky; 2008 but with less effects from boundary conditions). A 60% peak shows that the organization of working memory facilitates forward subsequent recalls. Even if we assume that the capacity of working memory makes the "effective" list length only 4 items, 60% is an overrepresentation: in a four item list there would be 12 different first recall pairs: item 1-item 2, item 1-item 3, etc. and only three (item1-item2, item2-item3 and item3-item4) that are forward subsequent, a 25% probability, which means that the 60% peak is more than twice as high as should be; in this same limited capacity case the backward subsequent recall is suppressed by a factor of two. The halfway point between the two stages is around conditional recall 3 (which is actual recall 4) which would make 4 a second estimate of the capacity of working memory.

Fourth, we can remove as much of the effect of the boundary conditions as possible by calculating the relative conditional response probabilities: the ratio between the real conditional response probabilities in Fig. 5 and the random values of Fig. 4.  The result is displayed in Fig. 6 in logarithmic plots. The forward subsequent peak remains (a huge 30 times chance) and the probabilities that make up this peak are taken from backward recalls for the first stage.  These are also the recalls for which previous items are the most strongly suppressed (suppressed by 26 times chance). Even in the second stage, there is a strong tendency for forward subsequent recalls (5 times chance).  The halfway point between the two stages is around conditional recall 3.5 which would make 4.5 a third estimate of the capacity of working memory.

Since the scale is so different for the different parts of Fig. 6, consider instead the integral of the differences between Fig. 5 and Fig. 4 in Fig. 7. The first stage shows a large transfer from the back items to the nearest neighboring items.  Subsequent recalls show a much smaller transfer from back and forward items into nearest neighboring items, which is consistent with Fig. 4 and the much larger delocalization of the second stage recalls.  Most of the suppressed recalls in the second stage seem to come from forward items.  The halfway point between the two stages is around conditional recall 4 which would make 5 a fourth estimate of the capacity of working memory.

Fig. 8 displays the relative conditional response probabilities from Fig. 6 as a function of recall for just the forward subsequent recall (left panel) and the average of the backward recalls from -40 to -10.  The two stages of recall can be seen in the large initial peak of the forward subsequent recall and the large initial valley of the backward recalls.  The halfway points are at the $3.5^{th}$ conditional recall (left panel) and $4^{th}$ conditional recall (right panel) yielding fifth and sixth estimates of working memory capacity of 4.5 and 5 word items.  The average of these estimates is 4.6

TABLES

| Work | Item types | List length and presentation interval | Recall interval | Subjects | Item presentation mode |
|---|---|---|---|---|---|
| *Murdock (1962)* | *Selection from 4000 most common English words, referred to as the Toronto Word Pool.* | *40 words in a list, each word presented once a second* | *1.5 minutes* | *15 undergraduates in data set* | *Verbal* |

*Table 1. Experimental method that generated the data used in this contribution.*

+

FIGURES

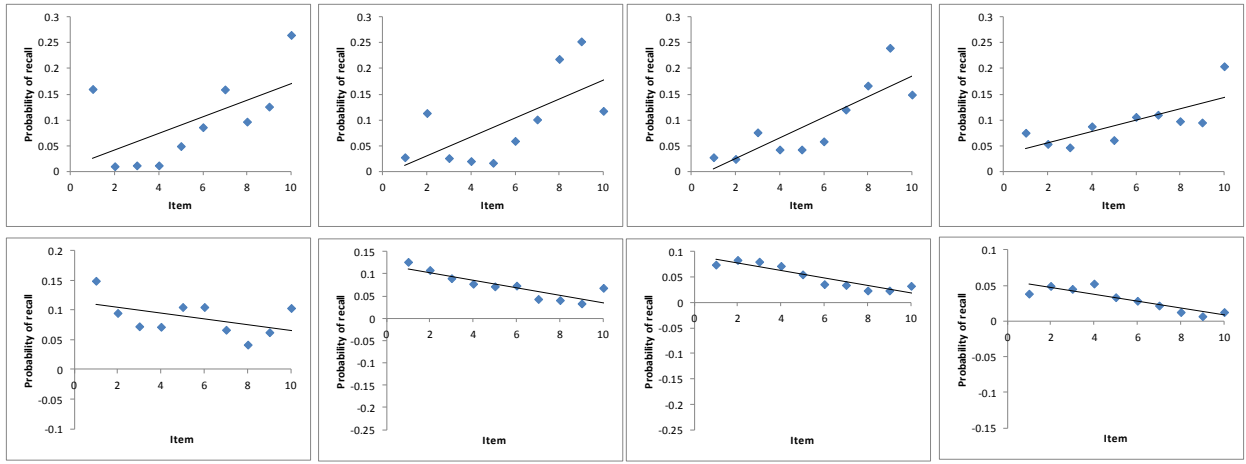

*Fig. 1. Recalls 1-4 (top panel) and 5-8 (bottom panel) for the 10-2 experiment of Murdock (1962). The first three recalls are from working memory, last three recalls from the second stage, and the 4$^{th}$ and 5$^{th}$ recalls are from a combination of working memory and second stage.*

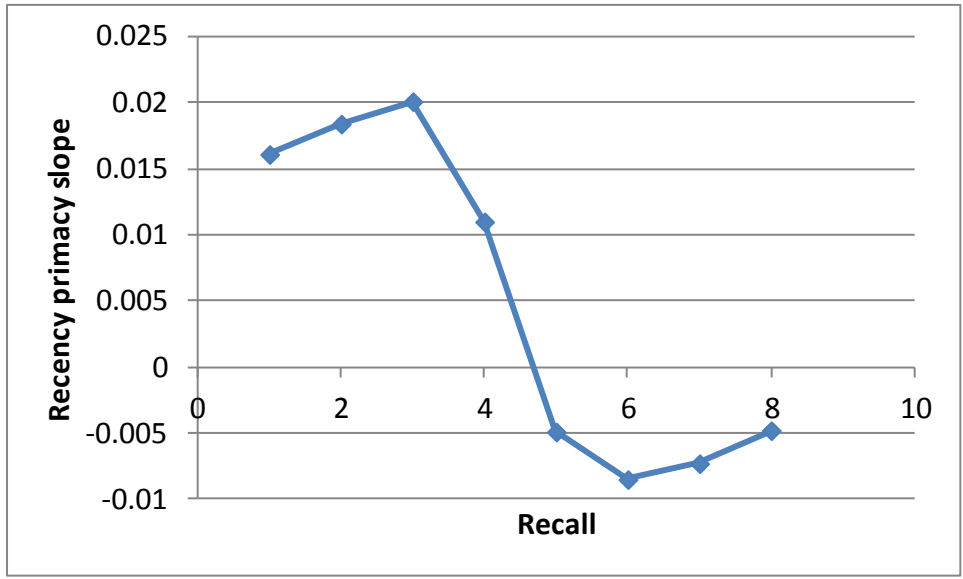

*Fig. 2. The slope of a linear fit to the serial position curves for the 10-2 data. Positive slope indicates recency, negative slope indicates primacy. Note the similarity to a step function. The middle of the step function is a little higher than 4, corresponding to the capacity of working memory.*

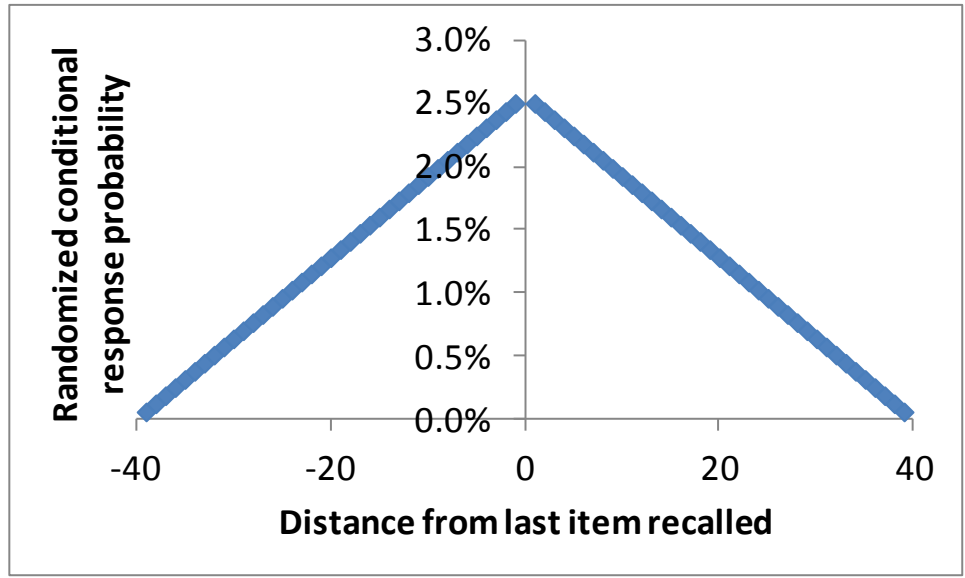

*Figure 3. Ideal conditional response probability curve if all recalls are equally likely and equally likely to be followed by any recall. For simplicity the condition that an item can only be recalled one has been ignored. Note the triangular shape which is a direct consequence of the finiteness of the list. The height of the triangle is 1/N where N is the list length - the longer the list the lower is the triangle – and 2.5% for the 40-1 dataset.*

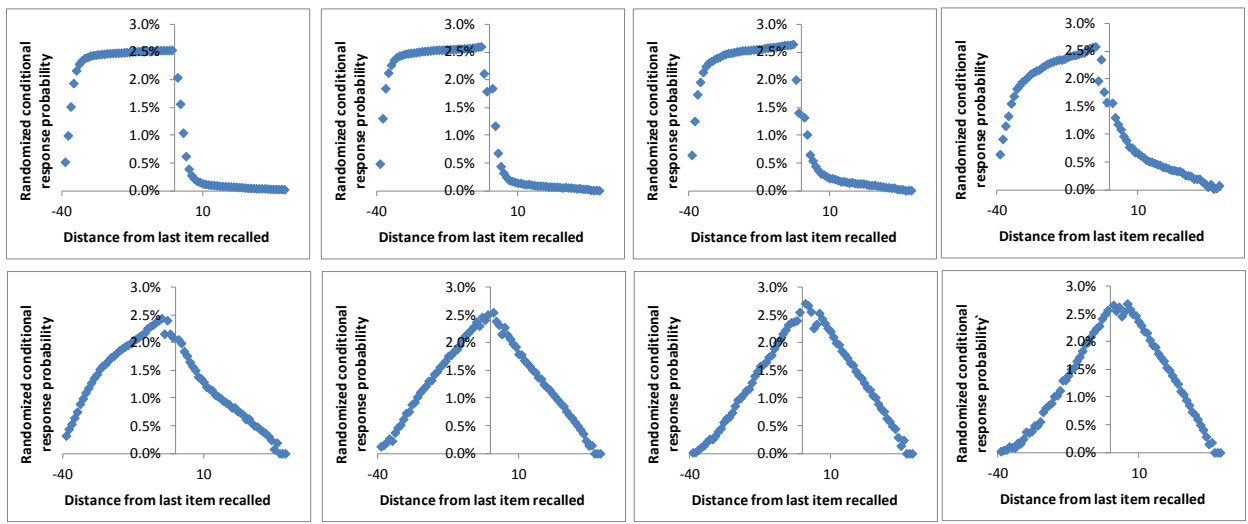

*Fig. 4. Conditional response probabilities after recalls 1-4 (upper panel) and 5-8 (lower panel) given recall history up to then and a subsequent random guess. The peaks measure about 2.5%. The first stage occurs after recalls 1-3 and the second stage after recalls 6-8.*

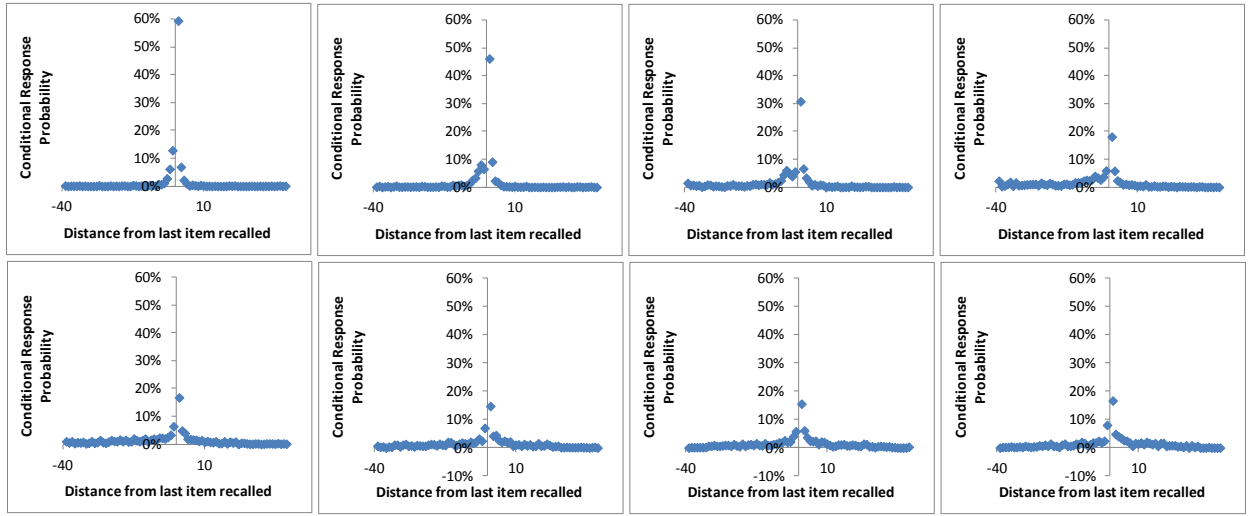

*Fig. 5. Actual conditional response probabilities after recalls 1-4 (upper panel) and 5-8 (lower panel). The peak starts out at 60% for the stage in which working memory is emptied and is gradually lowered to 15% for the second recall stage.*

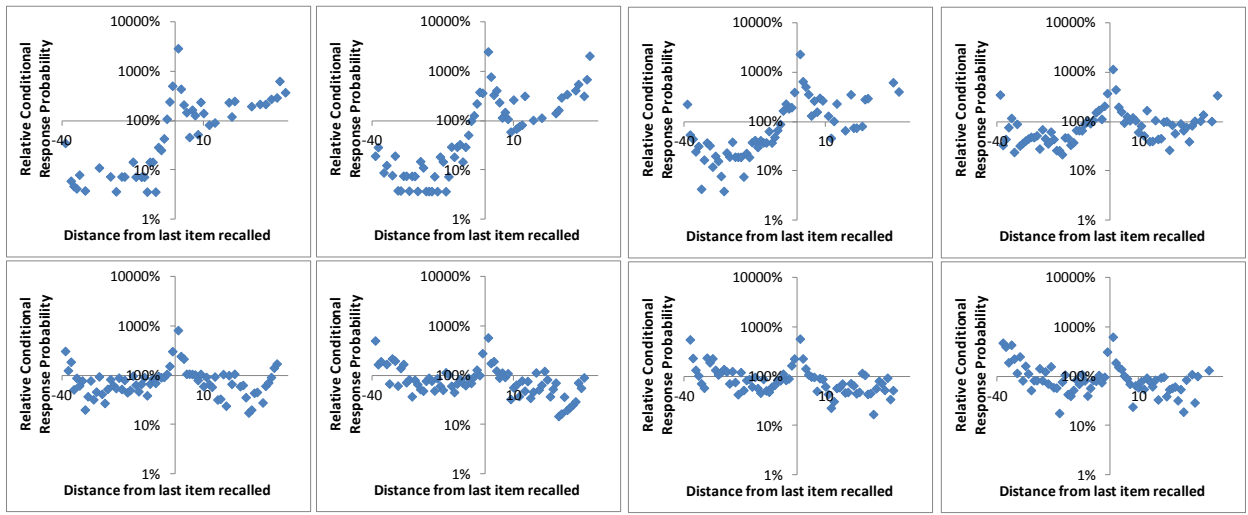

*Fig. 6. Ratio of conditional response probabilities after recalls 1-4 (upper panel) and 5-8 (lower panel) to the random conditional response probabilities of Fig. 4. The first stage occurs for conditional recalls 1-3 and the second stage for conditional recalls 5-8.*

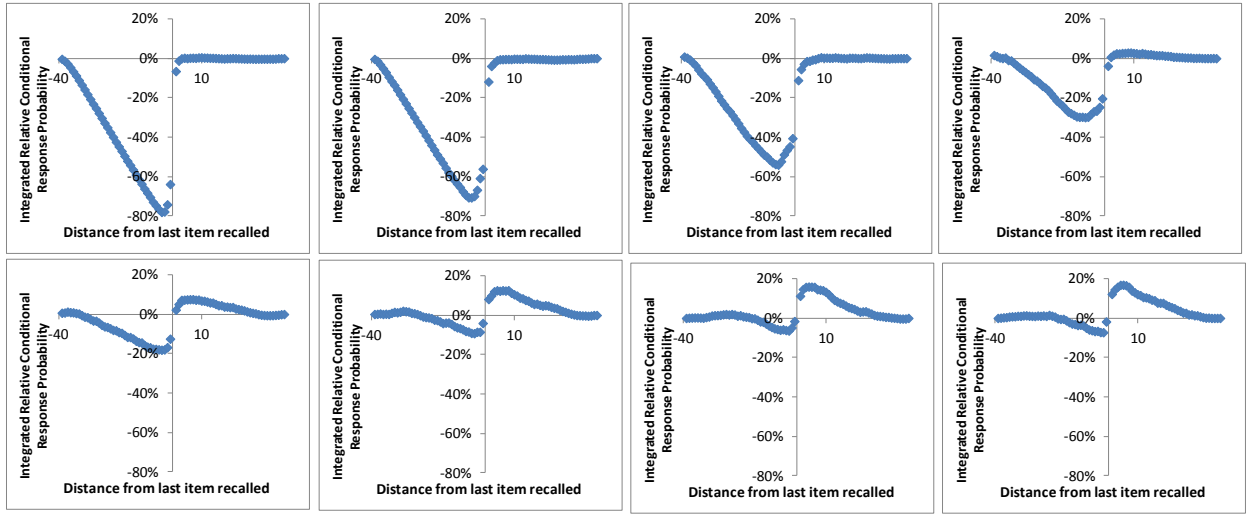

*Fig. 7. Integrated differences in conditional response probabilities (difference taken between the data in Fig. 5 and Fig. 4). First stage occurs for conditional recalls 1-3 and the second stage for conditional recalls 6-8.*

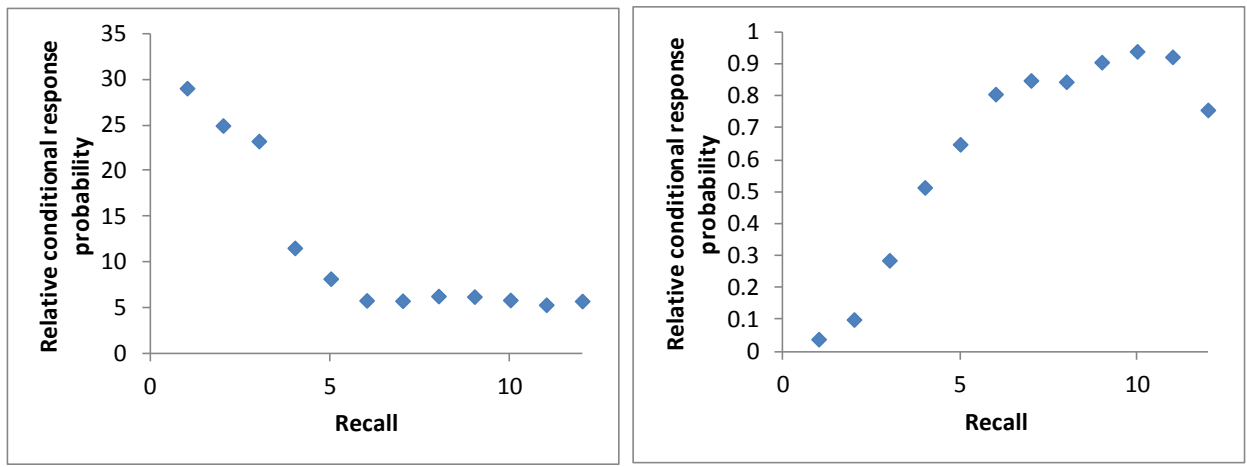

*Fig. 8. Relative conditional response probability as a function of recall for distance from last item = 1 (left panel) and ranging from -40 to -10 (right panel).  The initial enhancement is 30 in the left panel and the initial suppression is 26 in the right panel.*